\documentclass[epj]{svjour}

\usepackage{
graphicx
,epsfig
,ifthen
,amssymb
}
\usepackage{fancyhdr}
\def\makeheadbox{{
 }}
\newcommand{\beqn}{\begin{eqnarray}}
\newcommand{\eeqn}{\end{eqnarray}}
\newcommand{\nn}{\nonumber}
\newcommand{\ov}{\overline}
\newcommand{\<}{\langle}
\renewcommand{\>}{\rangle}
\newcommand{\D}{\Delta}
\newcommand{\id}{1 \hspace{1.15mm} \!\!\!\!1}

\def\mytitle{My title} 
\def\myauthors{My name}  
\def\mytype{My type of session}
\def\mysession{My session}

\def\mytitle{The MFV-MSSM: application to meson mixings} 
\def\myauthors{Diego Guadagnoli}    
\def\mytype{Contributed Talk}    
\def\mysession{Flavor Physics}

\begin{document}
\title{The Minimal Flavor Violating MSSM:\\application to meson mixings}
\author{
D.~Guadagnoli
\thanks{\emph{Email:} diego.guadagnoli@ph.tum.de}%
}                     
\institute{Physik-Department, Technische Universit\"at M\"unchen, D-85748 Garching, Germany}
%
\date{\today}
\abstract{
We provide a short account of the effects one expects on meson-antimeson oscillations
in the context of the Minimal Supersymmetric Standard Model (MSSM). This issue is 
largely dependent on the assumptions made on the MSSM parameter space. In this respect,
we consider in closer detail the cases of the general MSSM, with completely free soft
terms, and the Minimal Flavor Violating limit of the MSSM, providing a natural mechanism
of near-flavor-conservation. For the case of meson oscillations in the $\Delta B = 2$ sector,
we show that this approach leads to a striking increase of the predictivity of the model. 
In particular, we find
{\em (i)} SUSY corrections to be naturally small and always positive;
{\em (ii)} if $\mu$ is not small, an increase in importance  
(even for low $\tan \beta$) of scalar operators due to gluino  
contributions. The last point signals that (V$-$A)$\times$(V$-$A) dominated  
MFV is in general inconsistent with the MSSM. 
In this context, we also briefly discuss the MFV-Unitarity Triangle.
} 
\maketitle
\section{Introduction}\label{sec:intro}

Within the Standard Model (SM) flavor-changing neutral current (FCNC) effects are
forbidden at tree-level. Therefore FCNC observables are probes of the SM at the quantum
level, allowing in principle to identify the need for additional degrees of freedom circulating in the
loops. While the data on FCNC observables measured so far show no significant 
deviation from the SM expectations, in many extensions of the SM around
the electroweak (EW) scale predicted new effects tend to be visible if not dominating. 
In the case of the MSSM, this happens because of its soft sector: the ignorance of symmetries
underlying its structure compels to parameterize it most generally (general MSSM).
However, a completely general parameterization of the soft Lagrangian, 
besides impairing the predictive power of the model, turns out to imply also way too large 
FCNC effects in `most' \footnote{This notion of course depends on the metric used to explore the
parameter space.} of the MSSM parameter space. 

This puzzling circumstance is referred to as the `flavor problem', in SUSY made acute by the bulkiness 
of the soft sector parameter space. To visualize this problem, one can consider the
concrete FCNC example of $B_s - \bar B_s$ mixing. Within the general MSSM, one can focus
on the leading order contributions from the strong interacting sector, represented by
gluino-quark-squark vertices. Since flavor violation is
driven by the flavor off-diagonal entries of the (down-)squark mass matrix and the
relevant box diagrams feature two squark propagators, one expects the general structure:
SUSY correction $\sim \delta^2/M_{\rm SUSY}^2 \times f(\mbox{SUSY mass ratios})$, where
$\delta$ indicates a `mass-insertion', i.e. an off-diagonal entry in the
squark squared-mass matrix normalized to the geometric average of the corresponding two diagonal
entries. \footnote{The name `mass-insertions' is motivated by the fact that the squark
propagator is diagonalized perturbatively, by taking $\delta$'s as interactions.}
Since mixing measurements are, within errors, in agreement with the SM, the total
uncertainty -- by far dominated by the theoretical error on the lattice matrix elements, 
still exceeding 10\% -- can be translated into bounds on the $\delta$'s, or rather on 
$\delta/M_{\rm SUSY}$. Now, assuming $M_{\rm SUSY} =$ O(300 GeV) entails 
$|\delta| \lesssim 10^{-2} \div 10^{-3}$, which calls for an explanation in terms 
of symmetries \footnote{This problem is actually present
already in the quark mass matrices: they present disparate scales in the diagonal entries
and (before the rotation to the CKM basis) small off-diagonal entries.}. Assuming, 
on the other hand, $|\delta| =$ O(1) implies $M_{\rm SUSY} \gg$ O(TeV), posing again a problem 
of ``separation of scales''.

Two possible approaches to the SUSY flavor problem are the following: {\bf (i)} Focus on the
general MSSM and derive bounds on the $\delta$'s, however `fine-tuned' they may turn out
to be. Study effects allowed by these bounds on still to be measured observables; 
{\bf (ii)} Implement symmetry requirements on the soft terms and study their implications by 
exploring the more manageable parameter space resulting from the symmetry. The prototype of this 
kind of approach is Minimal Flavor Violation (MFV) \cite{MFV-USA,MFV}, in which FCNC effects in 
SUSY are small {\em because already those in the SM are}.

In the following two sections, I will provide just an example of the first approach
and then dwell more in detail on the second one.
\begin{figure*}[th!]
\centering
\includegraphics[width=0.33\textwidth]{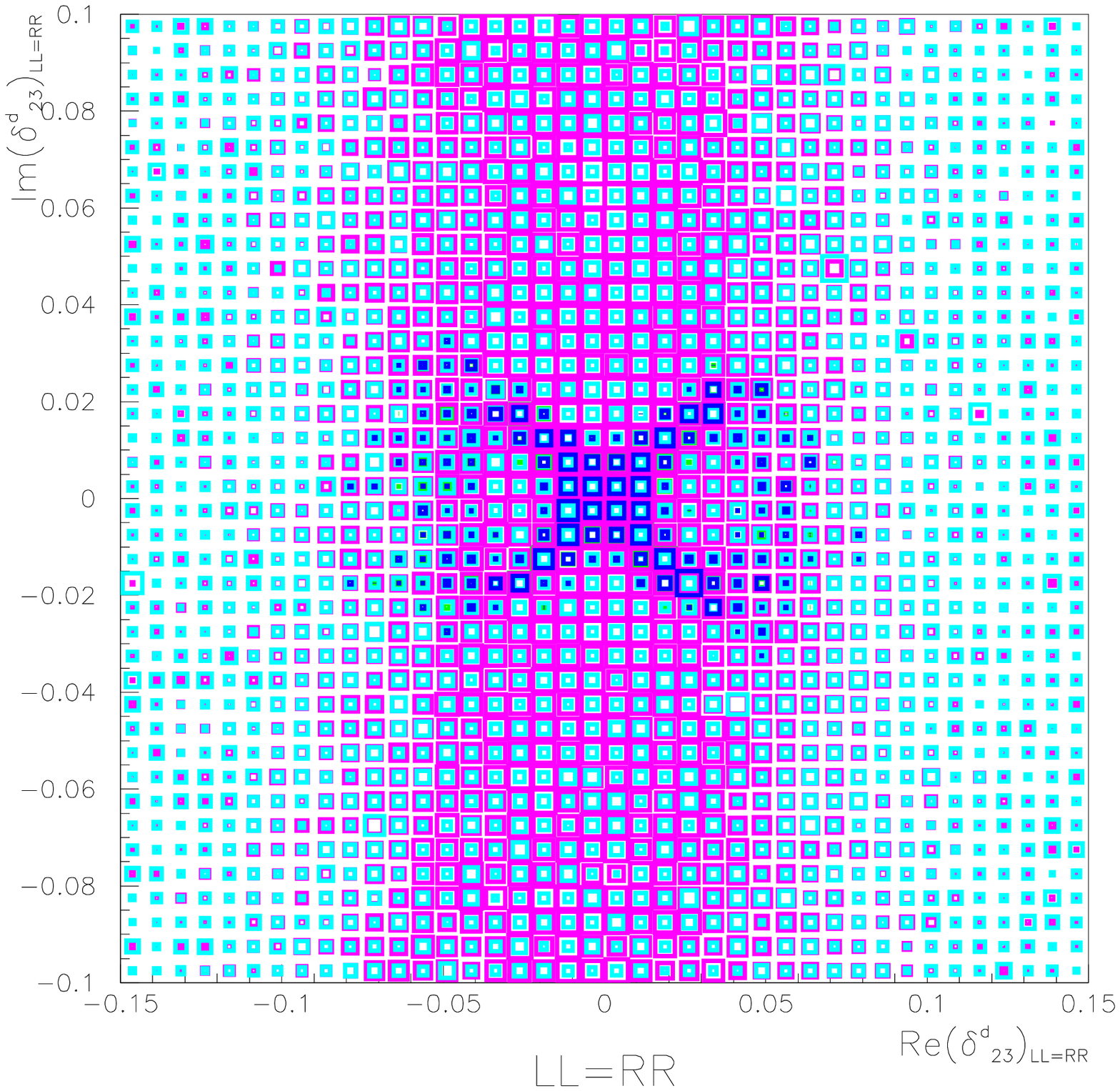}
\hspace{1.0cm}
\includegraphics[width=0.33\textwidth]{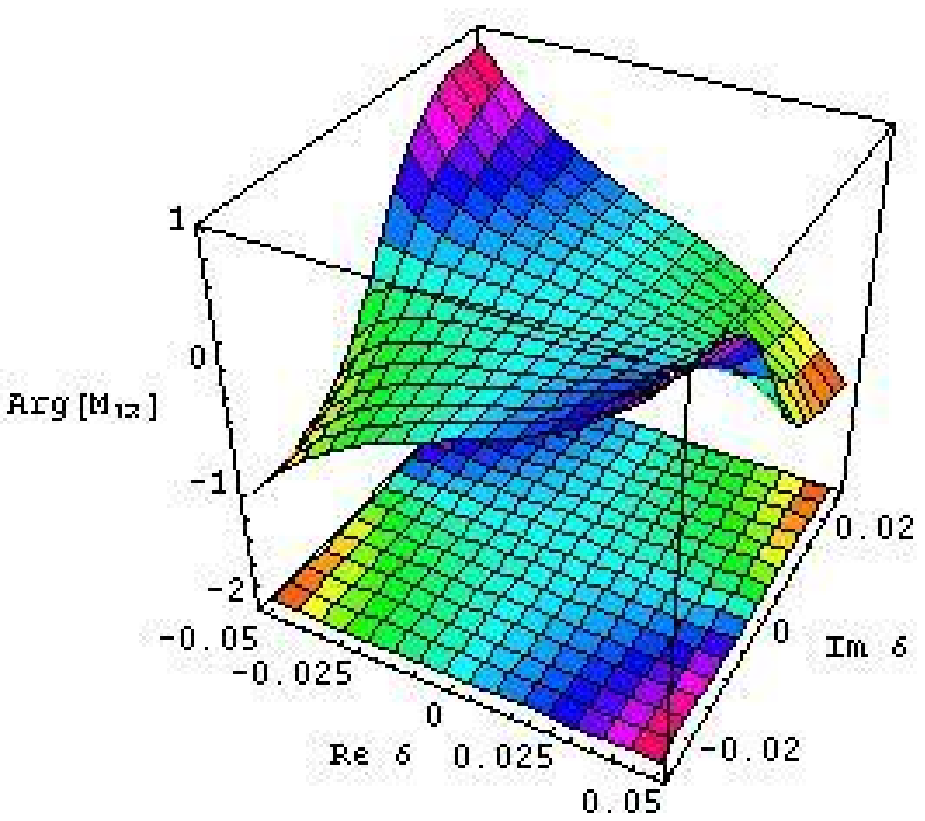}
\caption{Left: Constraints on ${\rm Re}(\delta^d_{23})_{LL=RR}$ for $\tan \beta = 3$,
$M_{\tilde q} = M_{\tilde g} = 350$ GeV. 
Constraints are from $\D M_s$ (green), $b \to s \gamma$ (pink), $b \to s \ell^+ \ell^-$
(cyan) and from the combined $b \to s$ transitions (blue). Right: $\arg(M_{12})$ profile
upon variation of $(\delta^d_{23})_{LL=RR}$ within the bounds obtained from the left
panel.}
\label{fig:dLLRR}
\end{figure*}

\section{General MSSM}\label{sec:gMSSM}

As an example, one can consider the bounds on the down-squark mass insertions imposed by
the most precise $b \to s$ transitions, namely $\D M_s$, $b \to s \gamma$ and 
$b \to s \ell^+ \ell^-$. In the general MSSM, one can limit oneself to gluino-mediated
contributions, and derive bounds on $(\delta^d_{23})_{AB}$, where 2,3 denote the external
flavors and $A,B = L, R$ are the superfield chiralities. Bounds on a given
$\delta$ are calculated by assuming its dominance, i.e. by setting all the other
$\delta$'s as zero. This approach is justified a posteriori by noting that the bounds on
different $\delta$'s are hierarchical. As an example, Fig. \ref{fig:dLLRR} (left) 
shows in a density plot the case $LL=RR$.
It is evident that the combined constraint implies the quite severe bound 
$|(\delta^d_{23})_{LL=RR}| \lesssim 5 \times 10^{-2}$.
One can now turn this bound into the maximum size of the predicted correction to
$\arg(M_{12})$, with $M_{12} \equiv \< \bar B_s | \mathcal H_{\rm eff}^{|\Delta B,S| = 2}|
B_s \>$. In the SM one has $\arg(M_{12}) \simeq 0.04$. In Fig. \ref{fig:dLLRR} (right) we
report the corresponding profile of $\arg(M_{12})$ within the MSSM, upon variation of 
$(\delta^d_{23})_{LL=RR}$ within the bounds obtained from the left panel of the same
figure. Notwithstanding the severe bound on the mass insertion, the phase of $B_s - \bar
B_s$ mixing can still be enhanced by up to two orders of magnitude with respect to the
tiny SM prediction. Access to this phase, via the measurement of the CP asymmetry in $B_s \to
\psi \phi$, will then provide a further, extremely powerful probe into $b \to s$ transitions.

\section{MFV-MSSM}\label{sec:MFV-MSSM}

From the example of the previous section, confronting the experimental data on $b \to s$
transitions with the corresponding predictions within the general MSSM, we are driven to the
conclusion that `generic' flavor violation, parameterized in terms of squark mass
insertions, has to be very small. If one rejects fine-tuning as an explanation, 
this fact calls of course for ungrasped symmetries, underlying the SUSY soft terms'
structure and implementing a mechanism of near-flavor-conservation \cite{MFV-USA}. 
In this respect, I turn now to discuss the case of the MFV limit of the MSSM. 
The starting observation is that, within the SM, FCNCs arise only because of the breaking of 
the flavor symmetry group due to the Yukawa couplings $Y_u, Y_d$ \cite{MFV-USA,MFV}. In
particular, the mechanism making FCNCs small within the SM is simply the specific
misalignment $Y_u$ and $Y_d$ entail between the quark flavor eigenbases and the corresponding
mass eigenbases. It is then interesting to address the question whether this specific
mechanism can also be {\em embedded} in extensions of the SM, where {\em new} flavor violating 
structures arise, a priori {\em unrelated} to the SM Yukawa couplings. The assumption that
the SM Yukawa couplings be, also in extensions of the SM, the {\em only} structures responsible 
for low-energy flavor and CP violation is known as MFV \cite{MFV}.
\footnote{This approach is more general than the so-called constrained MFV (CMFV)
\cite{BurasMFV,BBGT}, in which one also imposes the dominance of the SM operators. The 
phenomenological differences between MFV and CMFV have been spelled out in \cite{ABG}.} 
The MFV assumption implies that new sources of flavor violation become functions of the SM 
Yukawa couplings. In order to identify the functional dependence, Yukawa couplings are promoted to 
spurion fields of the flavor group \cite{MFV}. The resulting expansions for squark bilinear and 
trilinear soft terms are the following (for details on the formulae and on the notation see \cite{ABG})
\begin{figure*}[t]
\centering
\includegraphics[width=0.33\textwidth]{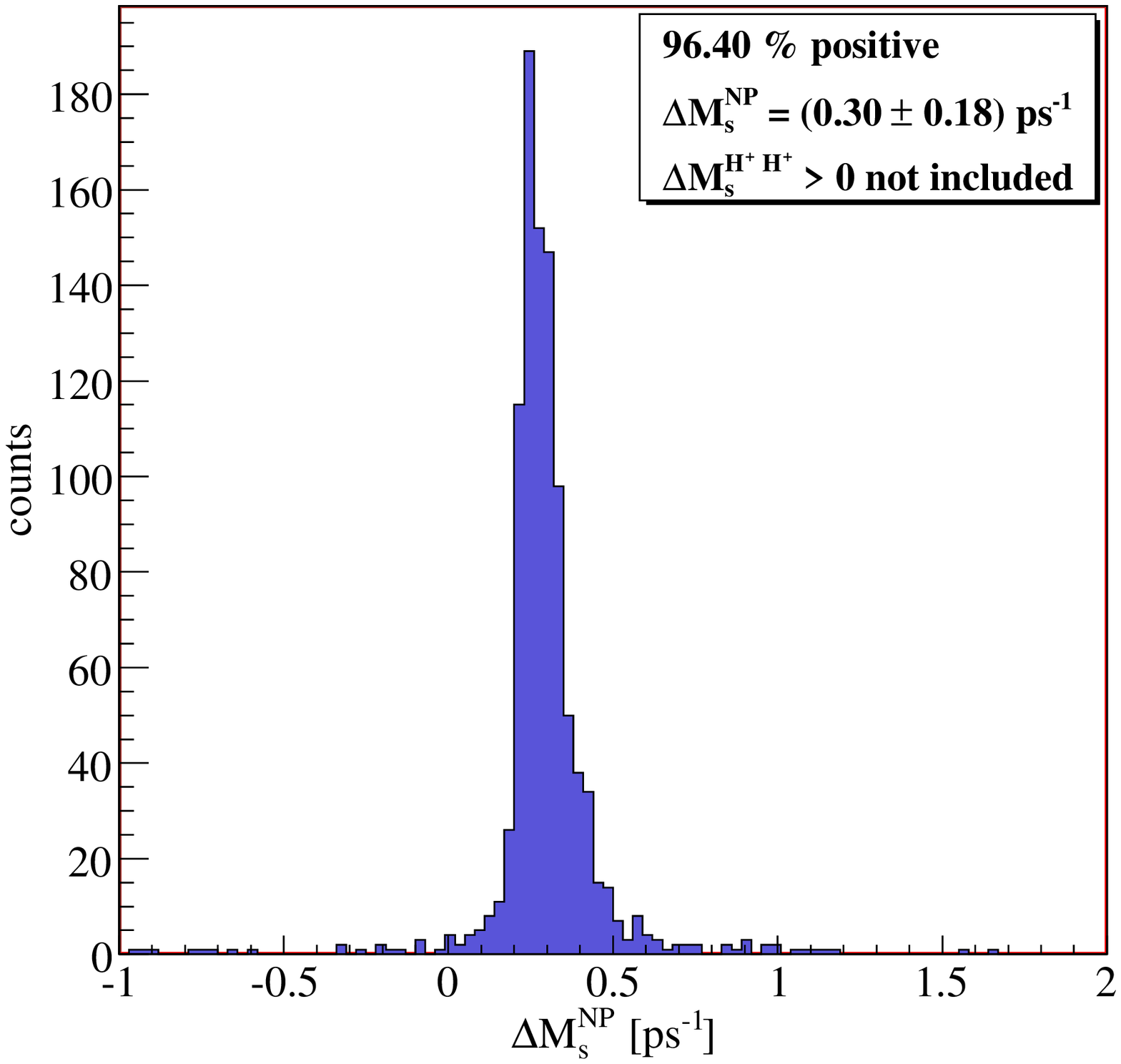}
\hspace{1.0cm}
\includegraphics[width=0.33\textwidth]{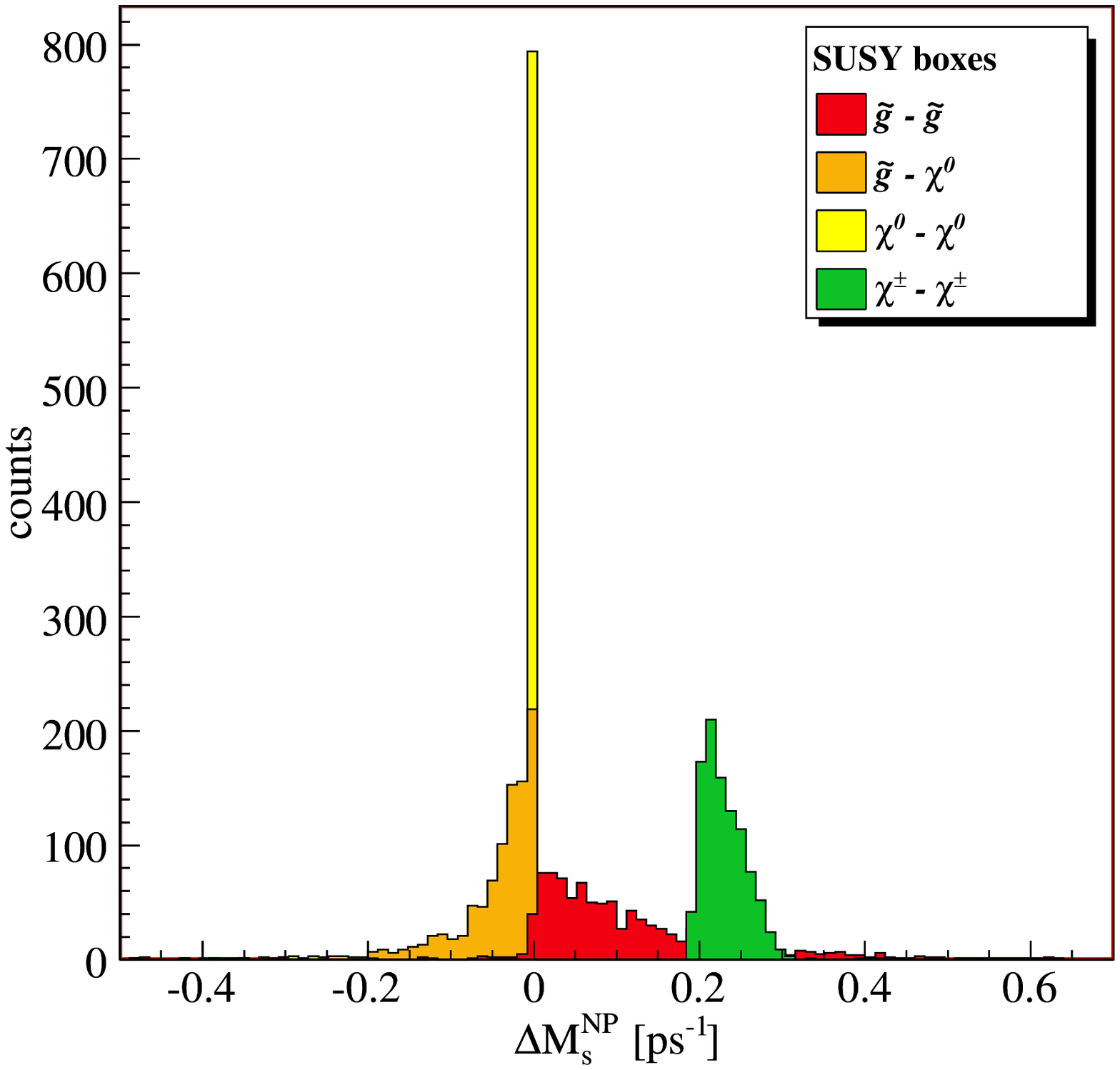}
%
%
\caption{Distribution of corrections to $\D M_s$ in the MFV-MSSM with $\tan \beta =
3$: sum of the SUSY contributions (left) and separate corrections (right). Mass scales are
chosen as (GeV): $\mu = 1000$, $\overline m = 300$, $M_{\tilde g} = 300$, $M_1 = 100$, $M_2 = 500$.}
\label{fig:DMs-MFV}
\end{figure*}
\beqn
[m_Q^2]^T &=& \ov m^2 \left( a_1 \id + b_1 Y_u Y_u^\dagger + b_2 Y_d Y_d^\dagger \right.\nn \\
&&\left. \phantom{x}\,+ b_3 (Y_d Y_d^\dagger Y_u Y_u^\dagger + Y_u Y_u^\dagger Y_d Y_d^\dagger) \right)~, \nn \\
m_U^2 &=& \ov m^2 \left( a_2 \id + b_4 Y_u^\dagger Y_u \right)~, \nn \\
m_D^2 &=& \ov m^2 \left( a_3 \id + b_5 Y_d^\dagger Y_d \right)~, \nn \\
A_u &=& A \left( a_4 Y_u + b_6 Y_d Y_d^\dagger Y_u \right)~, \nn \\
A_d &=& A \left( a_5 Y_d + b_7 Y_u Y_u^\dagger Y_d \right)~.
\label{softMFV}
\eeqn
As expansions (\ref{softMFV}) show, the assumption of MFV dramatically simplifies the
parametric dependence of the soft sector. In the case of meson mixings, to which we confine
our attention here, mass terms to be considered are the bilinear and trilinear mass scales 
$\ov m$ and $A$, respectively, the gluino mass $M_{\tilde g}$, the EW gaugino masses $M_{1,2}$, 
the $\mu$-parameter and the mass $M_{H^\pm}$ for the charged Higgs scalars
$H^\pm$. The remaining parametric dependence is on the (real) parameters $a_i, b_i$ ruling
the MFV expansions. 

Expansions (\ref{softMFV}) also explicitly show how, under the assumption of MFV, flavor
violating effects generated by the squark soft terms are {\em naturally small}. As an
example, after the rotation to the super-CKM basis, the flavor off-diagonal term $b_1 Y_u
Y_u^\dagger$ becomes $b_1 K^\dagger \hat Y_u^2 K$, with $K$ the CKM matrix and $\hat Y_u$
the diagonalized up-type Yukawa coupling. Considering $b_1$ an O(1) parameter, it is then
clear that, in this approach, the mass insertions of Section \ref{sec:gMSSM} become 
$\delta =$ O(1)$\times f(\mbox{CKM})$, showing the `CKM-like' nature of MFV effects in
SUSY.

A detailed study of meson mixings in the MFV-MSSM at low $\tan \beta$, adopting the above
approach \cite{MFV}, has been recently reported in Ref. \cite{ABG}. As already stated
above, the main purpose was there to spell out the differences in the approaches
\cite{MFV} versus \cite{BurasMFV,BBGT} to MFV, focusing on the benchmark case of meson mixings, 
where effects arising in MFV, and not reproducible in CMFV, are visible. \footnote{For an
interesting MFV study in an instance where effects beyond CMFV are however not visible, 
see Ref. \cite{IMPST}.} The strategy followed in this study was to fix mass scales to
``scenarios'' and, for each scenario, to study the distribution of corrections for 
$\D M_{d,s}$ assuming the $a_i, b_i$ coefficients (eq. (\ref{softMFV})) to be flatly
distributed in reasonable ranges (see \cite{ABG} for details).

The main features exhibited by the resulting distributions are as follows: 
{\em (i)} corrections are naturally small, typically not exceeding a few percent of the 
SM central value; 
{\em (ii)} corrections are typically spread in a narrow range for each mass scenario: the
standard deviation is smaller than the average correction; 
{\em (iii)} corrections are dominantly {\em positive}. This unexpected feature can then 
be considered a signature of the MFV-MSSM at low $\tan \beta$.

The above features are due to the interplay between chargino and gluino contributions,
while Higgs contributions do not depend on the $a_i, b_i$ and amount to just a further 
positive shift of the result, and neutralino contributions are always
negligible. The mass scales ruling this interplay are the squark mass scale $\ov m$ 
and the parameter $\mu$, with the other scales playing only a minor role. In particular, 
small values of $\mu$ with respect to $\ov m$ imply a small value for the lightest chargino 
mass and a correspondingly dominant (and positive \cite{gabrielli-giudice}) chargino
contribution. Conversely, large values of $\mu$, around 1 TeV, with a smaller squark scale, 
imply enhanced LR contributions in the squark mass matrix and correspondingly enhanced gluino
contributions. In addition, large values of $\mu$ also suppress the higgsino components of
chargino contributions. An example of this scenario is shown in Fig. \ref{fig:DMs-MFV}: here
chargino contributions are still dominant, but gluino contributions amount to positive 
corrections in the range $30 \div 50$\%, relative to charginos. From the right panel, it
is evident as well that gluino corrections also serve to compensate the negative contributions 
from gluino-neutralino boxes.

A final comment deserves the large $\tan \beta$ case. In this instance, it is well known
\cite{BCRSbig} that, even in the MFV-MSSM, large negative corrections 
to $\D M_s$ are possible, due to double Higgs penguins, enhancing the contributions from 
scalar operators. Since the latter are sensitive to the external quark masses, the same 
enhancement is typically negligible for the $\D M_d$ case. However, even in the $\D M_s$
case, allowed corrections turn out to be more limited when taking into account the new combined 
bound on the $B_s \to \mu^+ \mu^-$ decay mode from the CDF and D$\O$ collaborations
\cite{CDF+D0-bsmumu}. For positive $\mu$, corrections exceeding $-10$\% are basically
excluded. In order to still observe a relatively large effect on $\D M_s$, one needs negative 
values of $\mu$ and typically large values for $M_A \gtrsim 500$ GeV, increasing with increasing 
$\tan \beta \gtrsim 30$ \cite{ABGW}. One should however also keep in mind that for $\mu < 0$ the 
MSSM worsens the $(g-2)_{\mu}$ discrepancy with respect to the SM \cite{IP-MT}.

\section{The Unitarity Triangle in MFV}\label{sec:MFV-UT}
\begin{figure}[t]
\centering
\includegraphics[width=0.45\textwidth]{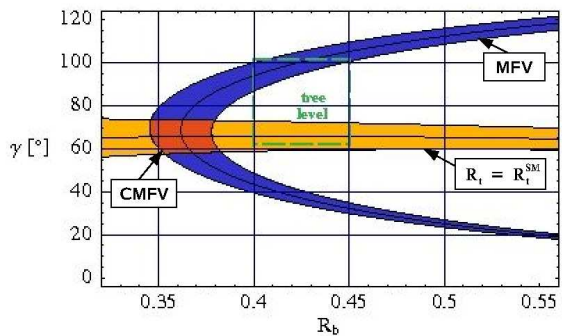}
\caption{Correlation between the UT angle $\gamma$ and the side $R_b$ in MFV models.
See text for details on the areas.}
\label{fig:gamma-Rb}
\end{figure}
Basing on the discussion in the previous section, the ratio $\D M_d/\D M_s$ is not
NP-independent in MFV, while it is equal to the SM ratio in CMFV, because of the
{\em assumed} dominance of the SM operator in the latter framework. The unitarity triangle (UT)
valid within CMFV is known as the Universal Unitarity Triangle (UUT) \cite{BurasMFV} and
is determined from the mentioned ratio $\D M_d/\D M_s$, allowing access to the side $R_t$, 
as well as from observables dominated by SM tree-level contributions and from angle measurements.
Among the tree-level dominated quantities one should mention in particular the value of 
$|V_{ub}|$, while the most precise angle determinations are $\beta$, measured by means of the
$S_{\psi K_S}$ asymmetry, and to a lesser extent $\gamma$. Extensive analyses of the UUT
have been performed by \cite{UTfit} (see e.g. \cite{UTfit-NP}) through global fits to the 
abovementioned quantities.

However, when assuming MFV, the UT should be determined exclusively from tree-level
observables and angle measurements. The side $R_t$ should instead not be included, since 
the MFV NP contributions to $\D M_d$ and $\D M_s$ do not generally cancel in the ratio.

In the context of MFV, the known value $\beta_{\psi K_S} = (21.2 \pm 1.0)^\circ$
\cite{HFAG} establishes a correlation \cite{ABG} between the side $R_b \propto |V_{ub}/V_{cb}|$
of the UT and the angle $\gamma$, that is valid for all models with MFV.
This correlation is represented in Fig. \ref{fig:gamma-Rb} as a blue area, under the 
assumption $\beta = \beta_{\psi K_S}$. The orange area in the figure
shows instead the region characterized by $R_t = R_t^{\rm SM}$, with an error dominated
by that of the lattice quantity $\xi = 1.23(6)$. The intersection between the two areas,
displayed in red, is then the one allowed to CMFV. Fig. \ref{fig:gamma-Rb} also shows 
the $1\sigma$-allowed range from tree-level decays, namely 
$64^\circ \le \gamma \le 102^\circ$ and $0.40 \le R_b \le 0.46$, as a green dashed box. The 
latter overlaps with the higher branch of the MFV area but not with the CMFV one. This
suggests that the slight tension \cite{BBGT} between the tree-level 
determination of $R_b$ and the one favoured by CMFV (which includes the SM) disappears within MFV, 
provided $\gamma \gtrsim 80^\circ$. One should however also keep in mind that this effect is
mainly driven by the discrepancy between the exclusive and the inclusive determinations of
$|V_{ub}|$ -- a not yet settled issue --, with the inclusive determination driving $R_b$ to somehow 
too high values with respect to those favored by indirect SM fits \cite{UTfit}.

\section*{Acknowledgments}
It is a pleasure to thank the organizers of SUSY 2007 for the proficient organization of the 
conference. I also warmly acknowledge interesting discussions with G.~Isi\-dori, S.~J\"ager,
O.~Vives and S.~Raby. Finally I warmly thank W.~Altmannshofer and A.~J.~Buras for many
useful comments on the manuscript and the A. von Humboldt Stiftung for support.

\end{document}